\documentclass{epl}

\title{Structure and Relaxation Dynamics of a Colloidal Gel}
\author{Emanuela Del Gado\inst{1} 
\and Walter Kob\inst{2}}
\institute{
\inst{1} Dip. di Scienze Fisiche, Universit\`a di Napoli "Federico II'',
80125 Napoli, Italy\\
\inst{2} Laboratoire des Collo\"\i des, Verres et Nanomat\'eriaux, Universit\'e Montpellier 2,
34095 Montpellier, France}
\begin{document}
\maketitle
\vspace*{-5mm}

\begin{abstract}
Using molecular dynamics computer simulations we investigate the structural
and dynamical properties of a simple model for a colloidal gel at low
volume fraction.  We find that at low $T$ the system is forming an
open percolating cluster, without any sign of a phase separation. The
nature of the relaxation dynamics strongly depends on the length 
scale/wave-vector considered and can be directly related to the geometrical
properties of the spanning cluster.
\vspace*{-4mm}
\end{abstract}
Gels are ubiquitous in daily life, biology, as well as technological
applications. Like glass-forming systems, they are disordered and have
slow and non-trivial relaxation ~\cite{ngai02,gel_dyn}. On the other hand,
they are characterized by open spanning structures~\cite{gel_stru} and hence
have a volume fraction $\phi$ that is significantly smaller than unity.
To what extent the structural properties are related to the dynamical
properties and what mechanism is responsible for the complex dynamics of
these systems, are important questions to which so far no clear answer
has been given. In particular, in view of the different nature of the
various gels (colloidal gels, chemical gels, etc.), it is not evident at
all that there are indeed unique answers.

In view of this variety it is not surprising that in the past various
mechanisms for the complex dynamics of gels have been proposed,
such as the occurrence of a percolation transition, the jamming
of preformed clusters, the arrested dynamics of a phase separating
system, and others~\cite{cates}-\cite{zaccarelli}.
Progress has also been hampered by the fact that there are so far very
few microscopic realistic models for gels that allow to investigate
these systems (in equilibrium!) by means of analytical methods or
computer simulations. In this letter we present 
a simple model where the gel formation is entirely due to the
interparticle interaction and that does indeed
have the characteristics of (colloidal) gel-forming systems at a finite
temperature. 
Subsequently we use the results on the structural and dynamical properties
of this system to shed some light on the mechanism that is responsible
for the slow dynamics in these systems.

The model investigated consists of identical particles of radius $\sigma$
that interact via an effective potential $V_{\rm eff}$ that is the
sum of a two-and three body terms. The two body potential is itself
the sum of a hard core like interaction that is given by a generalized
Lennard-Jones potential, $V_{\rm LJ}(r)$, and a term $V_{\rm cp}$ that
depends on the relative orientation of the particles. For the radial
term we have used 
$V_{\rm LJ}(r)= 23 \epsilon [(\sigma/r)^{18}-(\sigma/r)^{16}]$
where the prefactors and exponents have been chosen in such a way to
give a relatively narrow well of depth $\epsilon=1.0$ and width $0.2 \sigma$.
In the following we will measure length and energy in units such that $\sigma
=0.922$ and $\epsilon=1$, respectively, and time in units of $\sqrt{m
\sigma^2 /\epsilon}$, where $m$ is the mass of a particle. A system
with purely radial interaction is prone to undergo a phase 
separation~\cite{puertas}-\cite{zaccarelli}
and is therefore not a good model for a gel forming system. 
In fact, colloidal particles are seldom uniform and smooth and 
in particular, their roughness can produce quite 
rigid links between them~\cite{coll}. 
Actually the observation of equilibrium open 
structures \cite{gel_stru} strongly suggests an {\it effective} three body 
potential that favors the creation of an open network structure,  
as also recently measured~\cite{3b}.  In our case we have introduced
such an interaction by decorating each particle with 12 points that form
a rigid icosahedron of radius $1.1\sigma$.  The potential $V_{\rm cp}$
between a particle \#1 and a particle \#2 is then set up in such a way
that it is more favorable that the center of particle \#1 approaches
particle \#2 in the direction of one of the points of the icosahedron
that decorate particle \#2. In addition we have also included an explicit
(short range) three body potential $V_3$ in the form of a gaussian in the
angle $\theta$ between three neighboring particles and which makes that
values of $\theta$ smaller than $0.4 \pi$ are unlikely. More details
on these potentials will be given elsewhere~\cite{delgado_05c}. 
In the proposed model the (meso) particles can form directional bonds
that favor the formation of an open network structure, without, however,
imposing a local symmetry or connectivity, in contrast to models that
have been proposed before~\cite{delgado_latt,zaccarelli2}.

We have done microcanonical simulations
of this model using constrained molecular dynamics with the SHAKE algorithm~\cite{md} 
with a step size of 0.002. 
The number of particles was 8000 and the size of the simulation
box $L=43.09$ which gives a volume fraction of 0.05. (This corresponds
to a particle density of 0.1.)  Before starting the production runs
we carefully equilibrated the system by monitoring that the relevant
time correlation functions have attained their asymptotic 
limit~\cite{delgado_05c}. The
temperatures investigated were 5.0, 2.0, 1.0, 0.7, 0.5, 0.3, 0.2, 0.15,
0.1, 0.09, 0.08, 0.06, and 0.05. 
In order to improve the statistics of
the results we have averaged them over five independent runs.
\begin{figure}
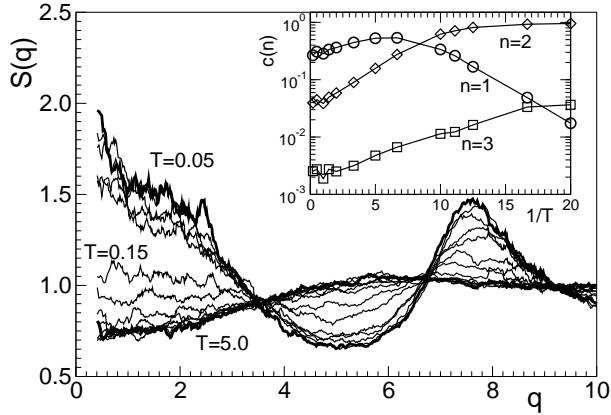

\onefigure[width=80mm]{fig1.eps}
\vspace*{-4mm}

\caption{
Main figure: Static structure factor $S(q)$ for all temperatures
investigated. Inset: $T-$dependence of $c(n)$, the fraction of particles
having coordination number $n$.}
\vspace*{-5mm}

\label{fig1}
\end{figure}

\begin{figure}
\onefigure[width=70mm]{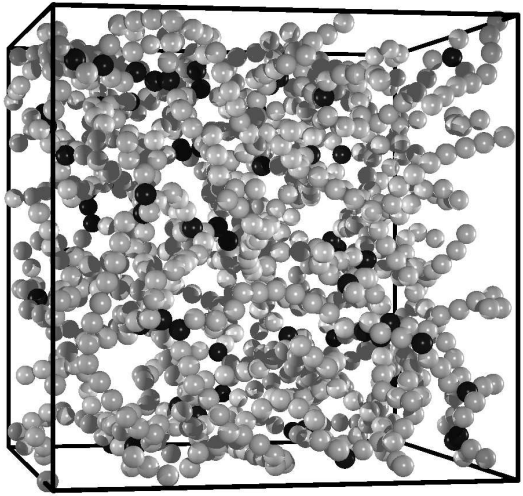}
\vspace*{-3mm}

\caption{
Part of a configuration, a cube with side $L/2$, of the system at $T=0.05$.
The light and dark particles have coordination
$n=2$ and $n=3$, respectively.}
\vspace*{-4mm}

\label{fig2}
\end{figure}

In Fig.~\ref{fig1} we show the static structure factor for all
temperatures investigated. We see that at high $T$ this function is
relatively flat, thus showing that the system has a structure that
is similar to the one of a gas of free particles. At around $T=0.1$
$S(q)$ starts to show a peak at $q_0=7.7$, a wave-vector which corresponds 
to the distance between two nearest neighbor particles. 
In addition $S(q)$ shows at small $q$ an increase which indicates the 
formation of a disordered open network structure. In fact this increase 
is relatively moderate and does not strongly depend on $T$, giving evidence 
that the system does not undergo a phase separation even at the 
lowest temperatures considered. Furthermore $S(q)$ does not show 
any pronounced peak at these low wave-vectors and thus we can conclude 
that the network is disordered and does not have a well defined length scale.
The appearance of the peak at $q_0=7.7$ is due to the fact that at
intermediate and low $T$ the particles condensate into clusters that
quickly grow and form at sufficiently low $T$ a percolating cluster. A
snapshot of a configuration at low $T$ is shown in Fig.~\ref{fig2}. This
figure clearly demonstrates that at low temperatures the typical
configurations are an open network in which most particles are forming
chains (light spheres) that are connected in a relatively random way
at points that have three or more neighbors (dark spheres). 

The change of the topology of the structure with decreasing $T$ can be
characterized by investigating the coordination number $c(n)$, which is
shown in the inset of Fig.~\ref{fig1}. We define $c(n)$ as the fraction of
particles that have exactly $n$ neighbors. (Two particles are considered
to be neighbors if their distance is less that $r_{\rm min}$=1.1, the
location of the first minimum in the radial distribution function.) We
see that at high temperatures the vast majority of the particles are
isolated, $n=0$ (not shown), and that the fraction of dimers, $n=1$,
is around 30\%. With decreasing $T$ this fraction increases, attains at
around $T=0.15$ a maximum, and then quickly decreases with decreasing
$T$. Thus we can conclude that at low $T$ the number of free particles
as well as the number of dimers (or chains ends) is very small. At the
same time the number of particles that have exactly two nearest neighbors
increases rapidly and these (local) 
configurations become by far the most prevalent
ones at low $T$. Last not least also the number of particles with $n=3$
neighbors increases quickly with decreasing $T$. From these curves we thus
can conclude that with decreasing $T$ the system forms an open network in
which most particles form chains that meet at points with coordination
number three and which are important for the mechanical properties
of the structure. Since on large length scales this structure is quite
homogeneous, $S(q)$ does not show a very pronounced increase 
at small $q$,
i.e. the compressibility of the system is relatively large but finite, in
agreement with the experimental results of gel-forming systems~
\cite{gel_stru}.
Note that the curve for coordination $n=3$ starts, at the lowest
temperature, to become flat, i.e. the number of particles that are
relevant for the size of the mesh of the network becomes independent
of $T$. From this we can conclude that, at low $T$, this mesh size,
and hence the overall structure, will only be a weak function of $T$,
in agreement with the result shown in Fig.~\ref{fig1} (main panel)
and experiments~\cite{gel_stru}. Of course it must be expected that
the asymptotic value of $c(n=3)$ depends on the concentration of the
particles in that the concentration of points of connectivity three
decreases with decreasing volume fraction.

\begin{figure}
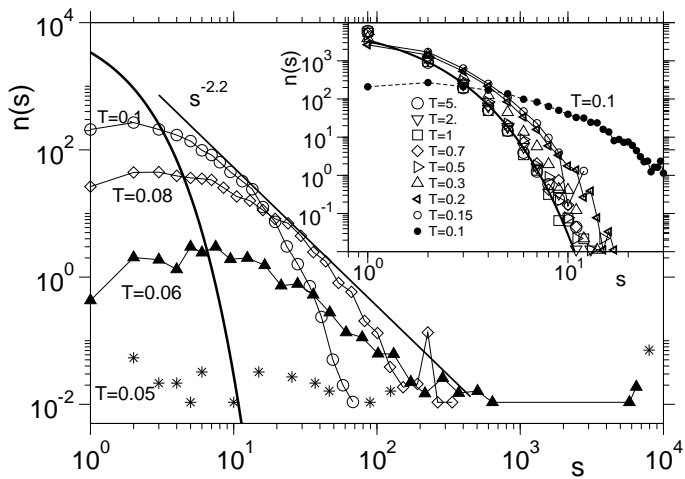

\onefigure[width=90mm]{fig3_n2.eps}
\vspace*{-3mm}

\caption{
Distribution of the clusters having size $s$ for all temperatures
investigated. Inset: $T\ge 0.1$. The bold line is an exponential
distribution. Main figure: $T \leq 0.1$. The bold line is the exponential
distribution from the inset. The straight line gives a power-law
distribution with exponent $-2.2$.}
\label{fig3}
\end{figure}
In order to characterize the structure of the system on a large length
scale it is useful to investigate $n(s)$, the number of clusters that
have exactly $s$ particles. (We define that a particle belongs to a
cluster if its distance from at least one member of the cluster is less
than $r_{\rm min}$.) This distribution is shown in Fig.~\ref{fig3}
for all temperature investigated. For $T\geq 0.3$ the distribution
nicely follows an exponential law (see inset), a behavior that corresponds to 
the random formation of transient clusters of non-bonded particles at low
densities. For the system size considered the largest of such clusters
found is around 15 particles, i.e. relatively small. At $T=0.1$ the
shape of the distribution has strongly changed in that now the most
probable clusters have size $s=2$. At the same time $n(s)$ has grown a
tail at large $s$ in that the largest clusters found have now $O(100)$
particles. At lower temperatures, $n(s)$ crosses over to a power law 
regime for high values of $s$, with a crossover point that moves to larger 
$s$ with decreasing $T$. At T=0.06 this regime is apparently compatible with 
an exponent around $-2.2$ (main plot in Fig.~\ref{fig3}), 
corresponding to random percolation~\cite{stauffer}.
At the lowest temperatures, $T\leq 0.06$, the
distribution shows a gap at large $s$ in that the system can form one
big cluster that contains a substantial fraction (more than 70\%) of
the particles in the system. Finally at $T=0.05$ we have only very few
particles that are members of small clusters, with an $n(s)$ that is
basically a constant, whereas the overwhelming majority of the particles
belongs to one large percolating cluster. Hence, at very low temperatures 
the distribution $n(s)$ is basically a constant since it is energetically 
very unfavorable to have small free clusters. 
(Note that the value of this ``constant'' decreases very rapidly with 
decreasing $T$, see Fig.~\ref{fig3}.)
These distributions show that for this system the formation of
the gel is intimately connected to the formation of a percolating
cluster \cite{de_candia,delgado_latt}.

\begin{figure}
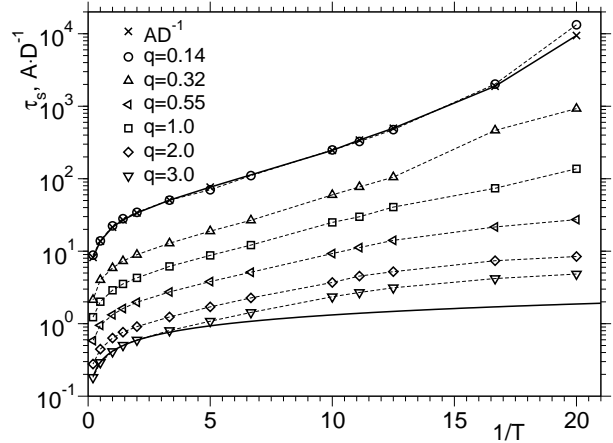

\onefigure[width=80mm]{fig4_n.eps}
\vspace*{-3mm}

\caption{
Arrhenius plot of the diffusion constant $D$ and of the relaxation
time $\tau_s(q,T)$ as determined from the self-intermediate scattering
function $F_s(q,t)$. The solid line is a fit to the high $T$
data for $q=3.0$ of the form $\tau={\rm const.}\cdot \sqrt{T}$.}
\vspace*{-3mm}

\label{fig4}
\end{figure}

Having characterized the structure of the system, we now turn our
attention to its dynamical properties. 
In view of the open network structure of the system at low temperatures,
it is of interest to investigate the lifetime of a bond between two
neighboring particles~\cite{delgado_05c} at intermediate and low temperatures.
For this we have determined $C_b(t)$, the
probability that a bond that exists at time zero is still present at time
$t$. We have found that the resulting $C_b(t)$ are well described
by an exponential with a relaxation time that shows an Arrhenius dependence.
At low $T$ ($T<0.1$) this is the longest relaxation time in the system.
As in Fig.\ref{fig2} we make the distinction of bonds that are connected to 
particles that have a coordination number of three 
(let's call them ``anchor particles'')
from the bonds that are connected to particles that have a coordination
number of two (``bridging particles''), and find that the lifetime of the
former bonds is around a factor of 20 smaller than the one of the latter, but
the $T-dependence$ is the same. This result is reasonable since the
anchor particles experience on average more mechanical stress than the
bridging particles and, due to their higher coordination, have less
possibilities to yield to this stress. Hence it is more likely that their
bonds are broken. Therefore one can envision the relaxation process in this
system that the connecting branches of the network detach from the anchor
particles, the branch reorients and attaches itself to a different branch,
thus creating a new anchor point.
(However, the possibility that a branch breaks at a bridging
particle can not be neglected completely since there are significantly
more bridging particles than anchor particles. Thus this type of motion
will contribute to the relaxation dynamics as well.)

In order to understand the role of the diffusive behavior of the particles, 
we have determined
the mean-squared displacement of a tagged particle, $\langle r^2(t)
\rangle=\langle |{\bf r}_j(t)- {\bf r}_j(0) |^2 \rangle$, which for high $T$ 
shows at early times a ballistic behavior and at long times a diffusive one 
~\cite{delgado_05c}.
From this function we used the Einstein relation to obtain the 
diffusion constant $D(T)$. Furthermore we have calculated the self-intermediate scattering
function $F_s(q,t)$ for wave-vector $q$, $
F_s(q,t) = N^{-1} \sum_{j=1}^N 
\langle \exp[i {\bf q} \cdot ({\bf r}_j(t)- {\bf r}_j(0))]\rangle$.
The discussion of the quite complex $q$ and $T$ dependence of $\langle
r^2(t) \rangle$ and $F_s(q,t)$, characteristic for a gel-forming
system, will be presented elsewhere~\cite{delgado_05c} and here
we focus only on the $T-$dependence of $D$ and the relaxation time
$\tau$. The latter quantity has been obtained by calculating the area
under $F_s(q,t)$. Figure~\ref{fig4} is an Arrhenius plot of $\tau_s(q,T)$
for intermediate and small wave-vectors. 
From
the figure we recognize that at high $T$ the relaxation time follows for 
all $q$ a $T^{-0.5}-$dependence which can be easily understood from the 
ballistic
motion of the free particles and the small clusters. (Recall that in this
$T-$range the distribution of the cluster size is independent of $T$,
see Fig.~\ref{fig3}). For wave-vectors that are large, $q\geq 2.0$,
the $T-$dependence of $\tau$ at low temperatures is relatively weak
but still somewhat stronger than $T^{-0.5}$. This can be understood
by realizing that on the corresponding length scales the particles can
still undergo an (almost) ballistic motion, despite the fact that they
are, at low $T$, connected to other particles, since the whole local
structure is moving ballistically. This type of motion is no longer
possible if one considers length scales that become comparable to the
size of the mesh of the network which is around 10 and thus corresponds
to a $q$ smaller than 1.0. For $q=0.55$ we find at low temperatures
a $T-$dependence of $\tau$ that is close to an Arrhenius law whereas
for smaller wave-vectors we find an even stronger $T-$dependence. On
length scales that are thus comparable or larger than the typical mesh
size of the network the $T-$dependence is thus very similar to the one
characterizing the slow dynamics of dense glasses. Finally we mention that 
wave-vectors around or larger than $q_0$ are not very relevant for the
slow dynamics of this system~\cite{puertas}, in strong contrast to usual
glass-forming systems that are at much higher volume fraction.

Also included in Fig.~\ref{fig4} is the inverse of the diffusion
constant $D$, scaled by a factor $A=49.2$ in order to make
it coincide with the value of $\tau_s(q=0.14,T=1.0)$. We find that in
the whole $T-$range investigated this quantity follows very closely
the $T-$dependence of $\tau_s(q,T)$ for small $q$, which is evidence
that in this system the relaxation of the structure is closely linked
to the diffusive motion of the particles, in contrast to the behavior
found in dense glass-forming systems in which 
the $T-$dependence of $D$ is weaker than the one of
$\tau_s$~\cite{richert02}. This is due to the fact that here at the 
lowest temperatures nearly all the particles belong to the network, 
as discussed above. It is also 
interesting to notice that at low temperatures 
the mean-squared displacement $\langle r^2(t)\rangle$ 
shows at intermediate times 
a small shoulder which is the signature of a (weak) caging effect with a 
localization length around 0.2~\cite{delgado_05c}, that is similar to the value 
found in dense glasses. In contrast to these systems we find that at later times
$\langle r^2(t)\rangle$ shows a pronounced {\it second} plateau with
a localization length around 10, a value which is comparable to the
mesh size. Thus we see that in this system the effective cages for the
particles are significantly larger than the interparticle distances,
a characteristics of a ``soft matter'' system.

In summary we have presented a new realistic model for gel forming
systems.  In particular this model has the advantage to allow the
investigation of such systems even at very low volume fractions {\it and}
in equilibrium.  The presented results show that the dynamics become
already glassy before the phase separation starts, i.e. 
the occurrence of the gel is not necessarily connected to the presence 
of an impeded phase separation~\cite{zaccarelli,manley}. 
Here the gel is related to the formation of ramified clusters that grow with 
decreasing $T$ and therefore connect to each
other to form a percolating network in which the anchor points are
connected by flexible branches. Since these clusters form and dissociate
quite easily, it is not clear whether a description of the gel-forming
process in terms of a jamming of clusters~\cite{cates} is appropriate,
since the clusters completely integrate into the network and thus loose
their identity faster than the time scale of the
relaxation time of the system.

It is evident that the anchor points are very important for the
mechanical stability of the structure. On length scales of the order of
the interparticle distance the structure is quite flexible whereas on
the scale of the mesh size of the network, which is given by the typical
distance between the anchor points, the structure becomes somewhat more
rigid due to the constrains imposed by the enhanced connectivity. However,
even on that length scale the system is relatively soft as compared to the
one found in dense glasses.  With decreasing temperature the connecting
chains will become stiffer and hence give rise to an increased effective
interaction between the anchor points.  
The mode-coupling theory of the glass transition~\cite{gotze99} proposes a 
mechanism that is able to rationalize why a small change in the 
coupling strength, in our case the effective interaction between the 
anchor points due to the connecting chains, gives rise to a very strong 
change in the relaxation time and non-exponential relaxation.  
Note that in our case the mechanism which produces the coupling 
enhancement of the effective interaction is due to the presence of the 
percolating open network, which connects the particles over different 
length scales. 
Thus it is reasonable to assume that the
super-Arrhenius dependence of $\tau_s(q,T)$ for wave-vectors corresponding
to the length scale of the mesh size, see Fig.~\ref{fig4}, is related
to the increased coupling between the anchor points, an interaction
that is transmitted by the connecting chains of the network through its 
structure, and that this mechanism is also the explanation for the strongly 
non-exponential relaxation dynamics~\cite{delgado_05c}.

Last not least we mention that since the lifetime of the bonds shows
only an Arrhenius dependence, it can be expected that at sufficiently
low temperatures the relaxation times of the system will also show an
Arrhenius dependence. This suggests that the system does not show
a real dynamical arrest at any finite temperature, in agreement with
the experimental findings~\cite{gel_dyn} and in strong contrast to the
more dense glasses for which an {\it apparent} dynamical singularity is
found at a finite temperature.
\vspace*{-4mm}

\acknowledgments 

Part of this work has been supported by the Marie Curie Fellowship
MCFI-2002-00573, the European Community's Human Potential Programme
under contract HPRN-CT-2002-00307, DYGLAGEMEM, and EU Network Number 
MRTN-CT-2003-504712.

\end{document}